\documentclass[aps,prd,preprint,floatfix]{revtex4}
\usepackage{epsfig}
\usepackage{bm}
\usepackage{dcolumn}
\usepackage{latexsym}

\begin{document}

\preprint{\rightline{ANL-HEP-PR-08-31}}

\title{Separating the scales of confinement and chiral-symmetry breaking
in lattice QCD with fundamental quarks}

\author{D.~K.~Sinclair}
\affiliation{HEP Division and Joint Theory Institute, Argonne National 
Laboratory, 9700 South Cass Avenue, Argonne, IL 60439, USA}

\begin{abstract}
Suggested holographic duals of QCD, based on AdS/CFT duality, predict that one
should be able to vary the scales of colour confinement and chiral-symmetry
breaking independently. Furthermore they suggest that such independent
variation of scales can be achieved by the inclusion of extra 4-fermion
interactions in QCD. We simulate lattice QCD with such extra 4-fermion terms
at finite temperatures and show that for strong enough 4-fermion couplings the
deconfinement transition occurs at a lower temperature than the
chiral-symmetry restoration transition. Moreover the separation of these
transitions depends on the size of the 4-fermion coupling, confirming the
predictions from the proposed holographic dual of QCD.
\end{abstract}

\maketitle

\section{Introduction}

The AdS/CFT duality of Maldacena \cite{Maldacena:1997re,Aharony:1999ti}, 
which indicates that ${\cal N}=4$
supersymmetric Yang-Mills field theory in 3+1~dimensions is dual to
${\cal N}=8$ super-gravity in 4+1~dimensions, inspired people to search for a
gravity/string theory which is dual to QCD. Since such duality would map the
strong-coupling regime of QCD to the weak-coupling regime of a classical
gravity theory, this promises to greatly simplify certain QCD calculations.
For a review of the early work in trying to construct such holographic duals
for QCD without quarks, we refer the reader to the review article by Aharony
\cite{Aharony:2002up}.
The inclusion of quarks was addressed by Karch and Katz \cite{Karch:2002sh}. 
Another such model for including quark flavours was suggested by Sakai and 
Sugimoto \cite{Sakai:2004cn,Sakai:2005yt}
and studied further by Antonyan, Harvey, Jensen and Kutasov 
\cite{Antonyan:2006vw} and Aharony, Sonnenschein and Yankielowicz 
\cite{Aharony:2006da}. 
These papers (\cite{Antonyan:2006vw,Aharony:2006da})
observe that the scales of confinement and chiral-symmetry
breaking can be varied independently in these proposed QCD duals, and that the
chiral-symmetry breaking scale must always be shorter than or equal to the
confinement scale. Note that all these attempts at constructing holographic
duals of QCD only claim validity for large $N_c$.

Measurement of the scales of confinement and chiral symmetry breaking can be
addressed in lattice QCD by measuring the deconfinement and chiral-symmetry
restoration temperatures for hot QCD. For quarks in the fundamental 
representation of $SU(3)$ colour, these two transitions have been observed to be
coincident, within the limitations of lattice measurements 
\cite{Polonyi:1984zt}. For this reason
there have been a number of attempts to explain why this should be so
\cite{Digal:2000ar,Mocsy:2003qw,Fukushima:2003fm,Hatta:2003ga}.
Others have argued that the scales of confinement and chiral-symmetry breaking
could easily be different \cite{Gross:1991pk}. 
However, it has been observed that for quarks in the adjoint representation
\cite{Engels:2005te},
and probably for quarks in the sextet representation of colour 
\cite{Kogut:1984sb}, 
the deconfinement temperature is significantly lower than the chiral-symmetry
restoration temperature. This suggests that the real reason why these
transitions appear coincident for fundamental quarks, is that the interaction
between fundamental quarks and antiquarks is too weak to produce a chiral
condensate at distances shorter than the confinement scale. The condensate
is then produced at the confinement scale since confinement requires 
chiral-symmetry breaking \cite{Banks:1979yr,Leutwyler:1992yt}. 
This would then make the two scales and hence the two transition temperatures
identical.

The work of Antonyan, Harvey, Jensen and Kutasov \cite{Antonyan:2006vw}
suggests that the introduction
of extra 4-fermion interactions in QCD would allow the separation of the
confinement and chiral-symmetry breaking scales. This makes sense, since it is
known that models of the Gross-Neveu \cite{Gross:1974jv}/Nambu-Jona-Lasinio 
\cite{Nambu:1961tp,Nambu:1961fr} type exhibit 
chiral-symmetry breaking without confinement for strong enough coupling. Hence
one might expect that the addition of a large enough 4-fermion coupling to QCD
could produce chiral-symmetry breaking without confinement. We therefore choose
to study lattice QCD with extra 4-fermion interactions of the
Gross-Neveu/Nambu-Jona-Lasinio type at finite temperatures, looking for 
evidence for separate deconfinement and chiral-symmetry restoration transitions.
Here we have been careful to make sure that the 4-fermion coupling is not so
strong as to produce spontaneous chiral-symmetry breaking without gauge fields,
since this phase is separated from the chiral-symmetry restored phase by a
bulk transition.

Early studies of such a model showed some evidence for separate transitions for
strong enough 4-fermion interactions \cite{Kogut:1998rg}, 
but the lattice size ($8^3 \times 4$)
was so small that the effect could have been a finite lattice size artifact.
We study this model for large 4-fermion couplings on $16^3 \times 4$ and
$24^3 \times 4$ lattices, and for intermediate 4-fermion couplings, on
$12^2 \times 24 \times 4$, $24^3 \times 4$ and $32^3 \times 4$ lattices (the  
$12^2 \times 24 \times 4$ `data' is from a previous study \cite{Kogut:2002rw}). 
For strong
4-fermion couplings, we find that the deconfinement transition takes place at 
a much lower temperature than the chiral-symmetry restoration phase transition.
At intermediate 4-fermion couplings, the 2 transitions are much closer, but
still clearly separate. At weak 4-fermion couplings, previous work has 
indicated that the two transitions are coincident within the limits of our
simulations \cite{Kogut:1998rg,Kogut:2002rw}.

In section~2 we introduce the action for lattice QCD with additional 4-fermion
interactions and discuss some of its properties, and how it is simulated. 
Section~3 presents our simulations at zero gauge coupling which are necessary
to identify the limits on the 4-fermion couplings we can use. In section~4 we
describe our simulations and present results. Section~5 discusses our results,
draws conclusions and indicates directions for future research.

\section{$\chi$QCD}

The lattice QCD action with additional 4-fermion interactions which we choose
is one we have previously called $\chi$QCD \cite{Kogut:1998rg}. 
This is based on the continuum Euclidean space-time Lagrangian density
\begin{equation}
{\cal L} = \frac{1}{4}F_{\mu\nu}F_{\mu\nu}+\bar{\psi}(D\!\!\!\!/+m)\psi
-{\lambda^2 \over 6 N_f}[(\bar{\psi}\psi)^2-(\bar{\psi}\gamma_5\tau_3\psi)^2].
\end{equation}
Introducing auxiliary fields $\sigma$ and $\pi$ yields a new equivalent
Lagrangian density which is quadratic in the fermion fields.
\begin{equation}
{\cal L} = \frac{1}{4}F_{\mu\nu}F_{\mu\nu}
         + \bar{\psi}(D\!\!\!\!/+\sigma+i\pi\gamma_5\tau_3+m)\psi
         + {3 N_f \over 2\lambda^2}(\sigma^2+\pi^2).
\end{equation}

The chosen 4-fermion term is of the Gross-Neveu/Nambu-Jona-Lasinio type. It
breaks flavour symmetry, but preserves the reduced chiral symmetry of the
staggered-fermion implementation of quarks on the lattice. Unlike the 
action which includes `$\tau_1$' and `$\tau_2$' terms and preserves the full
chiral flavour symmetry, it has a real positive fermion determinant which is
essential for lattice simulations. The terms involving the chiral auxiliary
fields ($\sigma$, $\pi$) remain local in the staggered-fermion lattice
transcription.

The lattice version of the action is
\begin{eqnarray}
S &=& \beta \sum_\Box \left[ 1-\frac{1}{3}{\rm Re}({\rm Tr}_\Box UUUU)\right]
     +\sum_{\tilde{s}}\frac{1}{8}N_f\gamma(\sigma^2+\pi^2) \nonumber \\
  && +\sum_{f=1}^{N_f/4}\sum_s\bar{\chi}_f\left[\not\!\! D+m+\frac{1}{16}\sum_i
               (\sigma_i+i\epsilon\pi_i)\right]\chi_f
\end{eqnarray}
where $\not\!\! D$ is the standard staggered gauge-covariant $\not\!\! D$,
$\epsilon=(-1)^{x+y+z+t}$, $s$ runs over the sites of the lattice, $\tilde{s}$
runs over the sites of the dual lattice on which the auxiliary fields reside
and $i$ runs over those sites of the dual lattice adjacent to the site of the
lattice on which the fermion field resides. This transcription preserves the
$U(1)$ chiral symmetry of the staggered fermion formulation. 
$\gamma=12/\lambda^2$ \footnote{The factor of 4 difference between this and
the identification in earlier papers is because there we followed earlier
work on the Nambu-Jona-Lasinio model, where $N_f$ was the number of staggered
fermion fields rather than the number of (continuum) flavours.}.

We simulate this action using the RHMC algorithm \cite{Clark:2006wp}, 
where the fractional powers
of the fermion determinant, required when the number of flavours $N_f$ is
not a multiple of 8, are obtained using a rational approximation to the 
fractional powers of the quadratic Dirac operator, and global Metropolis
accept/reject steps make the algorithm exact. For subtleties associated with
applying the RHMC algorithm to this action, see reference~\cite{Kogut:2006jg}. 
As noted in our
earlier work, the presence of the 4-fermion interaction makes the Dirac
operator non-singular in the chiral $m=0$ limit. We make use of this fact to
simulate at $m=0$, where there is an exact $U(1)$ chiral symmetry, so that the
chiral-symmetry restoration occurs at a true phase transition at and beyond
which the chiral condensate vanishes. There is, however, no true order
parameter known for the deconfinement transition in QCD with dynamical
fundamental quarks, which is seen as an abrupt jump in the Wilson Line
(Polyakov Loop). Hence this transition need not necessarily be a true phase
transition.

When we simulate at $m=0$, the direction in which the $U(1)$ chiral symmetry
is broken is arbitrary. The chiral condensate is some linear combination of
$\langle\bar{\psi}\psi\rangle$ and $i\langle\bar{\psi}\gamma_5\xi_5\psi\rangle$
or of $\sigma$ and $\pi$ ($\xi_5$, the flavour version of $\gamma_5$ is the
equivalent of $\tau_3$ for the 4 flavours described by a single staggered
quark field). On a finite lattice, the direction defined by this chiral
condensate rotates during the simulation, and the condensates average to zero
since there is no spontaneous symmetry breaking in a finite volume. We
therefore define an approximate order parameter, which approaches the true
order parameter when the volume of the lattice becomes infinite. Experience is
that $\sqrt{\bar{\psi}\psi^2-\bar{\psi}\gamma_5\xi_5\psi^2}$ or
$\sqrt{\sigma^2+\pi^2}$ are good choices. Here the quantities $\bar{\psi}\psi$,
etc. are lattice averages for a given gauge and auxiliary field configuration.
These choices lead to some finite size rounding of the phase transition on
finite lattices. From now on we shall denote these two versions of the chiral
condensate as ``$\langle\bar{\psi}\psi\rangle$'' and
``$\langle\sigma\rangle$''.

All our simulations have been performed with two flavours of massless quarks
($N_f=2$).

\section{Gross-Neveu/Nambu-Jona-Lasinio model simulations}

To determine the relevant values of $\gamma$, we perform simulations with the
pure Gross-Neveu/Nambu-Jona-Lasinio model, i.e. the theory described in the
previous section, only without gauge fields. For these simulations, we run at
zero temperature. Note that $N_f=2$ QCD with extra 4-fermion interactions
reduces to a 6-flavour Gross-Neveu/Nambu-Jona-Lasinio model (2 flavours 
$\times$ 3 colours) when the gauge coupling is set to zero. We are interested
in determining the $\gamma$ value for the bulk transition at which the chiral
condensate vanishes. Since, for our simulations with gauge fields, we want to
determine the two finite temperature transitions, we need to simulate with
4-fermion couplings weaker than that at the bulk transition, i.e. for 
$\gamma > \gamma_c$ where $\gamma_c$ is $\gamma$ at the bulk chiral-symmetry
restoration phase transition. In our earlier work, we simulated this theory
with only 4-fermion couplings, on an $8^4$ lattice. For this project we use a
$12^4$ lattice.

\begin{figure}[htb]
\epsfxsize=6in
\centerline{\epsffile{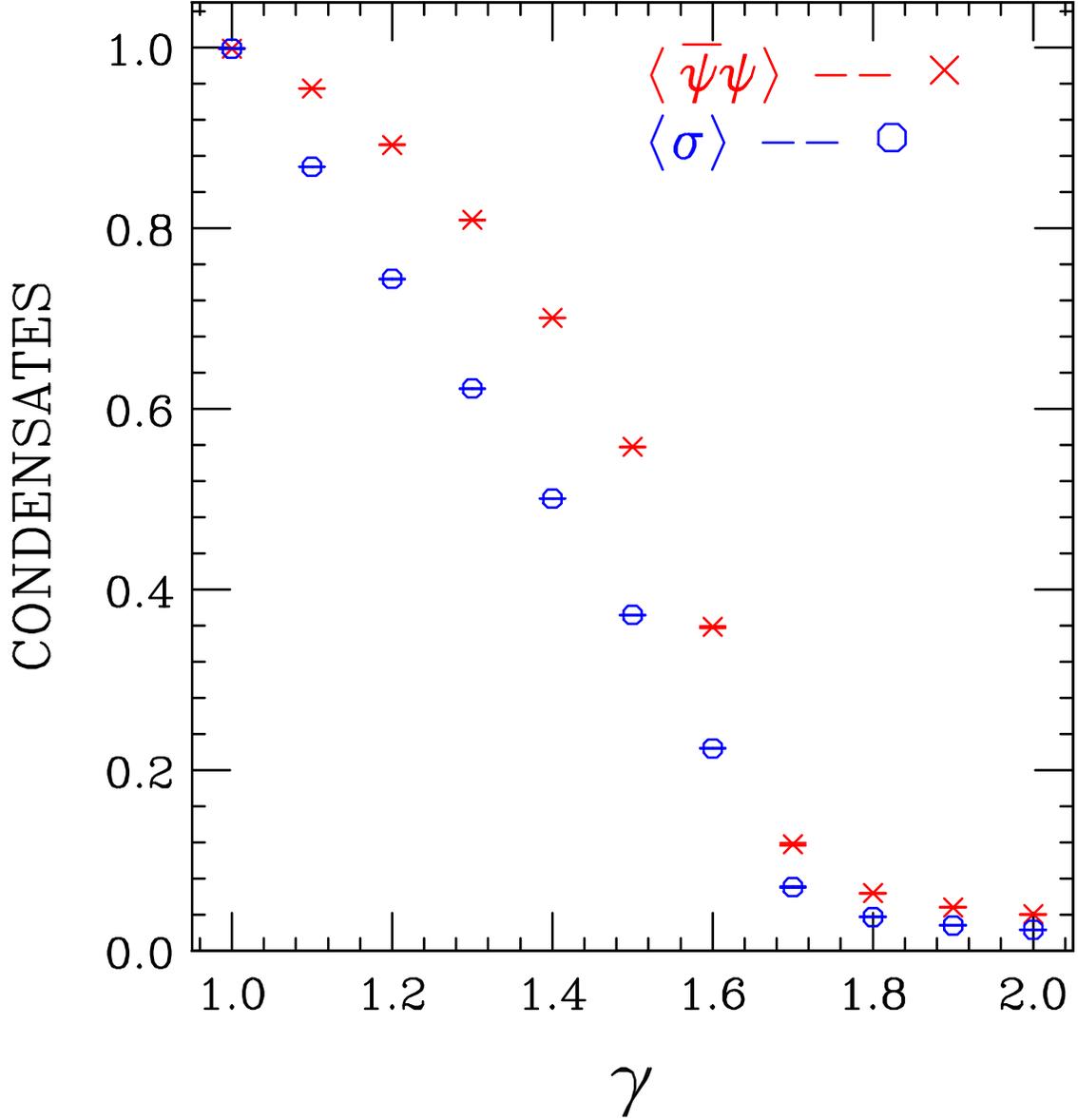}}
\caption{The chiral condensates $\langle\bar{\psi}\psi\rangle$, 
$\langle\sigma\rangle$ for the 6-flavour Gross-Neveu/Nambu-Jona-Lasinio (zero
gauge coupling) limit of QCD with extra 4-fermion couplings, as functions of
$\gamma$.}
\label{fig:njl}
\end{figure}

We performed runs for $1.0 \le \gamma \le 2.5$. For each $\gamma$ we generated
500,000 length 2.5 trajectories. Figure~\ref{fig:njl} shows the chiral
condensates $\langle\bar{\psi}\psi\rangle$ and $\langle\sigma\rangle$ as
functions of $\gamma$ over this range. From this we estimate $\gamma_c \approx
1.7$. Comparing this graph with that for an $8^4$ lattice from our previous
work, we observe that, as we approach $\gamma_c$ from below, the finite size
effects only become appreciable when we get very close to the transition, as
expected. The behaviour of $\langle\sigma\rangle$ can be understood from that
of $\langle\bar{\psi}\psi\rangle$, since in the limit of infinite volume, these
are related by
\begin{equation}
\langle\bar{\psi}\psi\rangle = \gamma \langle\sigma\rangle .
\end{equation}
This equation remains true when we reintroduce the gauge fields.

\section{Finite temperature simulations and results.}

We now turn to simulations with finite gauge couplings. The quark mass $m=0$
for all these simulations, so that the chiral transition is a true phase
transition. Since the separation of scales of confinement and chiral-symmetry
breaking is expected to occur for strong 4-fermion interactions, we start with
a large 4-fermion coupling. We choose $\gamma=2.5 > \gamma_c \approx 1.7$.
While this represents a large 4-fermion interaction it is not too close to the
bulk transition when the gauge coupling is taken to zero. Ideally one would
like to keep the gauge coupling fixed so that $\gamma$ is fixed in physical as
well as lattice units, changing the temperature by changing the temporal
extent (in lattice units), $N_t$, of the lattice. As is always the case (unless
one uses anisotropic lattices) this is impractical, and we fix $N_t$ and vary
the temperature by varying the lattice spacing in physical units by varying
$\beta=6/g^2$ where $g$ is the gauge coupling constant. Since we keep $\gamma$
fixed in lattice units, $\gamma$ changes in physical units as we vary $\beta$.
For these simulations we use $N_t=4$.

\begin{figure}[htb]
\epsfxsize=6in
\centerline{\epsffile{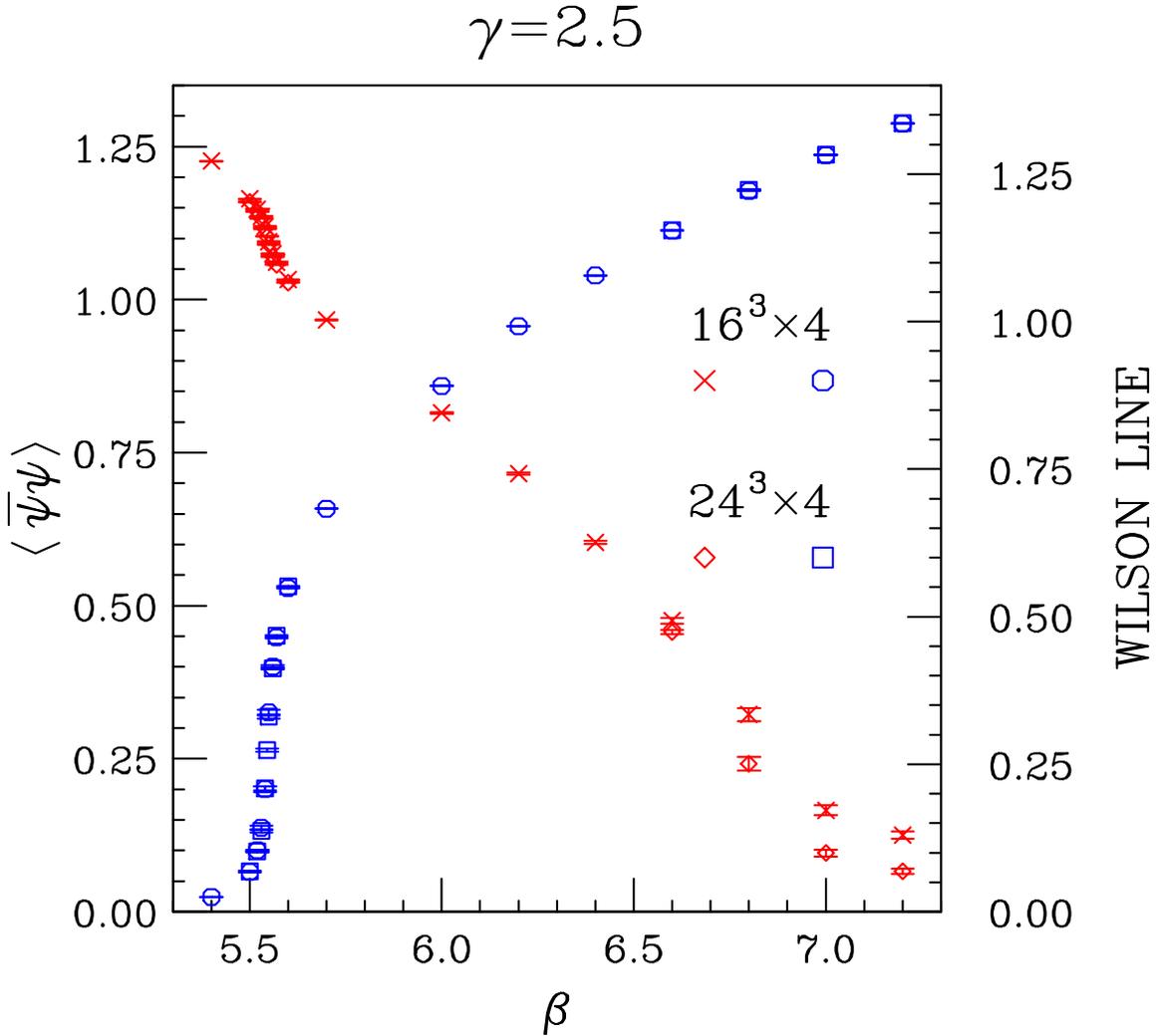}}
\caption{Wilson line and chiral condensate as functions of $\beta$ for 
$\gamma=2.5$ in lattice units.}
\label{fig:wil-psi2.5}
\end{figure}

Earlier simulations on $8^3 \times 4$ lattices at $\gamma=2.5$ suggested that
there could be two separate transitions. However, on such small lattices,
finite size effects are large enough that definite conclusions are suspect. We
have now simulated at $\gamma=2.5$ on $16^3 \times 4$ and $24^3 \times 4$
lattices. At each $\beta$ we ran for 50,000 length 1 trajectories, except for
$\beta=5.545$, close to the deconfinement transition, where we ran for 100,000
length 1 trajectories. In figure~\ref{fig:wil-psi2.5} we plot the Wilson Line
(Polyakov Loop), which signals the deconfinement transition and the chiral
condensate, which is the order parameter for chiral-symmetry breaking, against
$\beta$. The deconfinement transition, marked by a very rapid increase of the
Wilson line from near-zero, occurs at a much smaller $\beta$ than the chiral
phase transition, where the chiral condensate vanishes and chiral symmetry is
restored. In the region of the deconfinement transition, we use
Ferrenberg-Swendsen reweighting \cite{Ferrenberg:1988yz} to determine the
position of this transition from the Wilson line susceptibility peaks.
Examining the distribution of plaquette values for each $\beta$ in the range
$5.50 \le \beta \le 5.57$ suggests that the deconfinement $\beta$, $\beta_d
\approx 5.545$. Ferrenberg-Swendsen reweighting from $\beta=5.545$ yields
$\beta_d =5.547(3)$. The chiral transition is considerably less well measured,
since there are much larger gaps between consecutive $\beta$s in its
neighbourhood, and the finite size effects are large. Examining the `data'
from the two lattice sizes and trying to take into account the finite size
effects leads us to $\beta_\chi=7.0(2)$ as our estimate for the $\beta$ at the
chiral phase transition. Thus, for $\gamma=2.5$ the deconfinement and
chiral-symmetry restoration temperatures are far apart.

\begin{figure}[htb]
\epsfxsize=6in
\centerline{\epsffile{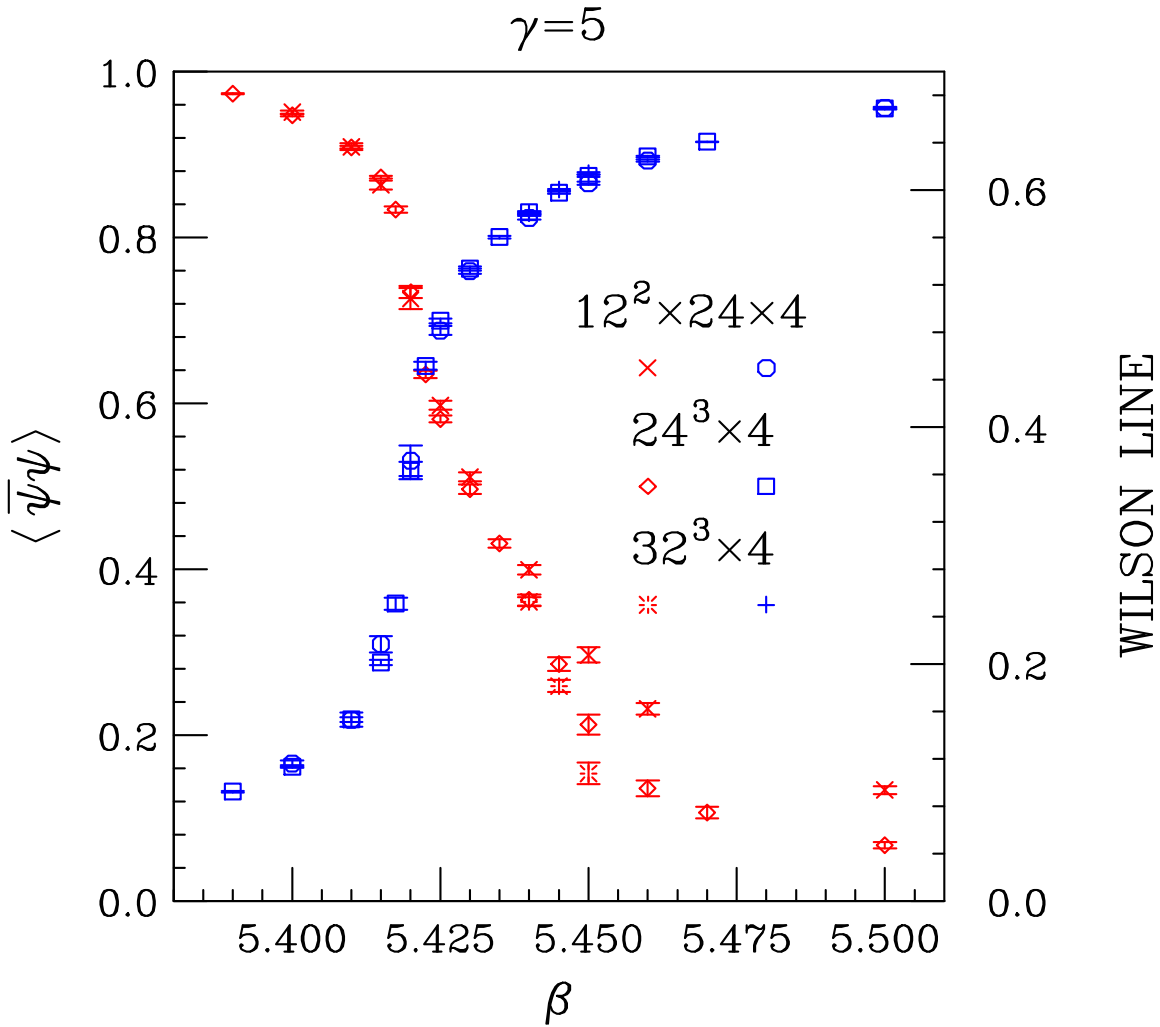}}
\caption{Wilson line and chiral condensate as functions of $\beta$ for
$\gamma=5$ in lattice units.}
\label{fig:wil-psi5}
\end{figure}

We also performed simulations at an intermediate 4-fermion coupling $\gamma=5$.
Here, earlier simulations on $8^3 \times 4$ and $12^2 \times 24 \times 4$
lattices had failed to indicate whether there was one transition or two.
We have performed simulations on $24^3 \times 4$ lattices. To clarify chiral
symmetry restoration, we also performed simulations on $32^3 \times 4$ lattices
close to the chiral-symmetry restoration transition. For the $24^3 \times 4$
lattice we have run for 100,000 length 1 trajectories for each $\beta$ in the
range $5.415 \le \beta \le 5.460$ and for 50,000 for those $\beta$s outside 
this range. On the $32^3 \times 4$ lattice we ran for 100,000 length 1
trajectories for each of three $\beta$ values. Figure~\ref{fig:wil-psi5}
shows the chiral condensate and Wilson line from these new simulations along
with the old results on $12^2 \times 24 \times 4$ lattices. Again we find
evidence for two separate transitions. Close to the deconfinement transition,
we find that the finite size effects are very small, as was observed at
$\gamma=2.5$. This is reassuring, especially in light of the fact that our
old $12^2 \times 24 \times 4$ simulations used the inexact R algorithm, while
the new simulations used the exact RHMC algorithm. We estimate that the 
deconfinement $\beta$, $\beta_d=5.420(4)$. Our estimated $\beta$ for the chiral
transition is $\beta_\chi=5.450(5)$.

We know from our earlier work that if $\gamma$ is increased to $\gamma=10$,
the deconfinement and chiral transitions appear to be coincident. The closeness
of $\beta_d$ and $\beta_\chi$ at $\gamma=5$ compared with $\gamma=2.5$
suggests that the two transitions come together at a $\gamma$, a little above
$5$.

\section{Discussion and Conclusions}

We have simulated lattice QCD with extra 4-fermion interactions of the
Gross-Neveu/Nambu-Jona-Lasinio type at finite temperatures, and have observed 
that, for strong enough 4-fermion couplings, the deconfinement and 
chiral-symmetry restoration temperatures are different. Increasing the 
4-fermion coupling increases the separation of these two transitions, and
the deconfinement temperature is always less than or equal to the 
chiral-symmetry restoration temperature. This agrees with the predictions
from proposed holographic duals of QCD \cite{Antonyan:2006vw,Aharony:2006da}.

These results can be easily understood, since Gross-Neveu/Nambu-Jona-Lasinio
models without gauge fields can exhibit spontaneous chiral-symmetry breaking,
but do not confine the fermions. It also indicates that the reason why for QCD
without such terms, the two transitions appear coincident is merely that the
eigenvalue of the quadratic Casimir operator for the fundamental
representation of $SU(3)_{colour}$ is too small for the QCD interactions to
produce chiral-symmetry breaking at distances less than the confinement scale.
Confinement then forces chiral symmetry breaking and the two scales are
identical.

When the two scales are different, chiral symmetry is broken on both sides of
the deconfinement transition. Hence, at this transition, the gauge fields
see `constituent quarks', i.e. quarks with dynamical masses produced by
spontaneous chiral-symmetry breaking, rather than the massless `current quarks'.
These are less effective at screening colour than the `current quarks' and the
deconfinement temperature increases. We expect that in the limit of large
4-fermion couplings the dynamical quark masses will increase without bound
and the deconfinement temperature will approach its quenched value. This
explains why the transition coupling $\beta_d$ and hence temperature $T_d$ vary 
over a limited range as the 4-fermion coupling is varied. These arguments 
suggest $5.25 \lesssim \beta_d \lesssim 5.6925$, where the lower limit is
from simulations without the 4-fermion interaction
\cite{Bernard:1999fv,D'Elia:2005bv} and the upper limit is the quenched value
\cite{Brown:1988qe,Bacilieri:1988yq}, over the whole range of $\gamma$ (the
inverse 4-fermion coupling), $\gamma_c < \gamma < \infty$, where 
$\gamma_c \approx 1.7$ is the bulk transition $\gamma$ when the gauge
couplings are switched off. On the other hand, the chiral-symmetry restoration
temperature $T_\chi$ and coupling $\beta_\chi$ will vary from $T_\chi=T_d$ and
$\beta_\chi=\beta_d$ for $\gamma > \gamma_0$ (weak 4-fermion coupling) to
$\infty$ as gamma is decreased to $\gamma=\gamma_c$, when chiral-symmetry is
always broken \footnote{Strictly speaking this occurs at 
$\gamma=\gamma_c(T) < \gamma_c(0)$ because, at $N_t=4$, the pure 4-fermion
theory is also at a finite temperature $T$.}. Our simulations indicate that
$\gamma_0$, the $\gamma$ value for which the two transitions coalesce, lies
between $5$ and $10$, and is probably closer to $5$.

We have simulated at $\gamma=2.5$ (strong 4-fermion coupling) and $\gamma=5$
(intermediate 4-fermion coupling) and previously at $\gamma=10$ and
$\gamma=20$ (both weak 4-fermion couplings)\cite{Kogut:2002rw}. The
deconfinement $\beta$s are $\beta_d(2.5)=5.547(3)$, $\beta_d(5)=5.420(4)$,
$\beta_d(10)=5.327(2)$ and $\beta_d(20)=5.289(1)$, consistent with the above
bounds, and increasing as expected with increasing 4-fermion coupling. The
$\beta$s for the chiral transition are $\beta_\chi(2.5)=7.0(2)$,
$\beta_\chi(5)=5.450(5)$, $\beta_\chi(10)=\beta_d(10)$ and
$\beta_\chi(20)=\beta_d(20)$. For $\gamma=2.5$ (strong 4-fermion coupling)
crude estimates based on quenched running of the coupling constant, from
earlier lattice simulations, indicate that $T_\chi$ is an order of magnitude
larger than $T_d$, while for intermediate 4-fermion coupling, $\gamma=5$,
2-flavour 2-loop running of the coupling constant yields 
$T_\chi \approx 1.04 T_d$.

In our model, the 4-fermion interaction is completely local and thus irrelevant
in the renormalization-group sense. Thus it will vanish in the continuum limit
when the lattice spacing goes to zero, and we will no longer have two separate
transitions. On the other hand, the 4-fermion interaction implied by the
proposed holographic dual to QCD is non-local, which softens ultra-violet
divergences introducing the possibility that it might have a non-trivial  
continuum limit. This point needs further investigation. Even if this does
not happen, and QCD with extra 4-fermion interactions is only defined with
an ultra-violet regulator such as the one provided by the lattice, it is a
useful model since it allows one to study confinement and chiral-symmetry
breaking independently in this ultra-violet regulated (effective) theory.

We have restricted ourselves to $N_t=4$ for this preliminary study. A more
complete study would require extending this to larger $N_t$. At $N_t=4$, we
would need more $\beta$ values to accurately pinpoint the chiral phase
transition and to determine its nature. Additional work would be needed to
understand the deconfinement transition. The fact that the vicinity of the
deconfinement transition shows little finite-size dependence, makes it likely 
that, if it is indeed a phase transition, then it is first-order.

A lattice analysis of the non-local Nambu-Jona-Lasinio model suggested by
the proposed string/gravity dual to QCD should be considered. It is a 
4-dimensional non-gauge theory which exhibits chiral-symmetry breaking but not 
confinement. If it has a non-trivial continuum limit, it represents a new
class of 4-dimensional field theories. The next step would be to include it
in the lattice QCD action, just as we have done with the local 
Gross-Neveu/Nambu-Jona-Lasinio model, or alternatively to study the 
5-dimensional gauge theory with right- and left-handed quarks pinned to 
separate 4-dimensional branes, which produced it.

The methods of lattice gauge theory and extensions of AdS/CFT duality to QCD
and similar quantum field theories can be used to complement one another in
the understanding of such theories. The particular example described in this
paper shows how these ideas can be applied to further the understanding of
confinement and chiral-symmetry breaking in QCD.

\section*{Acknowledgements}

We thank Jeffrey Harvey and David Kutasov for asking the questions which led to
these studies. We also acknowledge helpful discussions with Cosmas Zachos and
Alexander Velytsky. In addition, we thank John Kogut for permitting us to use
results from our previous work \cite{Kogut:2002rw}. This research was
supported in part by US Department of Energy contract DE-AC02-06CH11357, and
in part under a Joint Theory Institute grant. The simulations were performed
on NERSC's Cray XT4, Franklin. The results from the earlier work included in
figure~\ref{fig:wil-psi5} were obtained using the HP Superdome at the
University of Kentucky under an NRAC grant.


\begin{thebibliography}{999}

%\cite{Maldacena:1997re}
\bibitem{Maldacena:1997re}
  J.~M.~Maldacena,
  %``The large N limit of superconformal field theories and supergravity,''
  Adv.\ Theor.\ Math.\ Phys.\  {\bf 2}, 231 (1998)
  [Int.\ J.\ Theor.\ Phys.\  {\bf 38}, 1113 (1999)]
  [arXiv:hep-th/9711200].
  %%CITATION = IJTPB,38,1113;%%

%\cite{Aharony:1999ti}
\bibitem{Aharony:1999ti}
  O.~Aharony, S.~S.~Gubser, J.~M.~Maldacena, H.~Ooguri and Y.~Oz,
  %``Large N field theories, string theory and gravity,''
  Phys.\ Rept.\  {\bf 323}, 183 (2000)
  [arXiv:hep-th/9905111].
  %%CITATION = PRPLC,323,183;%%

%\cite{Aharony:2002up}
\bibitem{Aharony:2002up}
  O.~Aharony,
  %``The non-AdS/non-CFT correspondence, or three different paths to QCD,''
  arXiv:hep-th/0212193.
  %%CITATION = HEP-TH/0212193;%%

%\cite{Karch:2002sh}
\bibitem{Karch:2002sh}
  A.~Karch and E.~Katz,
  %``Adding flavor to AdS/CFT,''
  JHEP {\bf 0206}, 043 (2002)
  [arXiv:hep-th/0205236].
  %%CITATION = JHEPA,0206,043;%%

%\cite{Sakai:2004cn}
\bibitem{Sakai:2004cn}
  T.~Sakai and S.~Sugimoto,
  %``Low energy hadron physics in holographic QCD,''
  Prog.\ Theor.\ Phys.\  {\bf 113}, 843 (2005)
  [arXiv:hep-th/0412141].
  %%CITATION = PTPKA,113,843;%%

%\cite{Sakai:2005yt}
\bibitem{Sakai:2005yt}
  T.~Sakai and S.~Sugimoto,
  %``More on a holographic dual of QCD,''
  Prog.\ Theor.\ Phys.\  {\bf 114}, 1083 (2006)
  [arXiv:hep-th/0507073].
  %%CITATION = PTPKA,114,1083;%%

%\cite{Antonyan:2006vw}
\bibitem{Antonyan:2006vw}
  E.~Antonyan, J.~A.~Harvey, S.~Jensen and D.~Kutasov,
  %``NJL and QCD from string theory,''
  arXiv:hep-th/0604017.
  %%CITATION = HEP-TH/0604017;%%

%\cite{Aharony:2006da}
\bibitem{Aharony:2006da}
  O.~Aharony, J.~Sonnenschein and S.~Yankielowicz,
  %``A holographic model of deconfinement and chiral symmetry restoration,''
  Annals Phys.\  {\bf 322}, 1420 (2007)
  [arXiv:hep-th/0604161].
  %%CITATION = APNYA,322,1420;%%

%\cite{Polonyi:1984zt}
\bibitem{Polonyi:1984zt}
  J.~Polonyi, H.~W.~Wyld, J.~B.~Kogut, J.~Shigemitsu and D.~K.~Sinclair,
  %``Finite Temperature Phase Transitions In SU(3) Lattice Gauge Theory With
  %Dynamical, Light Fermions,''
  Phys.\ Rev.\ Lett.\  {\bf 53}, 644 (1984).
  %%CITATION = PRLTA,53,644;%%

%\cite{Digal:2000ar}
\bibitem{Digal:2000ar}
  S.~Digal, E.~Laermann and H.~Satz,
  %``Deconfinement through chiral symmetry restoration in two-flavour QCD,''
  Eur.\ Phys.\ J.\  C {\bf 18}, 583 (2001)
  [arXiv:hep-ph/0007175].
 %%CITATION = EPHJA,C18,583;%%

%\cite{Mocsy:2003qw}
\bibitem{Mocsy:2003qw}
  A.~Mocsy, F.~Sannino and K.~Tuominen,
  %``Confinement versus chiral symmetry,''
  Phys.\ Rev.\ Lett.\  {\bf 92}, 182302 (2004)
  [arXiv:hep-ph/0308135].
  %%CITATION = PRLTA,92,182302;%%

%\cite{Fukushima:2003fm}
\bibitem{Fukushima:2003fm}
  K.~Fukushima,
  %``Relation between the Polyakov loop and the chiral order parameter at
  %strong coupling,''
  Phys.\ Rev.\  D {\bf 68}, 045004 (2003)
  [arXiv:hep-ph/0303225].
  %%CITATION = PHRVA,D68,045004;%%

%\cite{Hatta:2003ga}
\bibitem{Hatta:2003ga}
  Y.~Hatta and K.~Fukushima,
  %``Linking the chiral and deconfinement phase transitions,''
  Phys.\ Rev.\  D {\bf 69}, 097502 (2004)
  [arXiv:hep-ph/0307068].
  %%CITATION = PHRVA,D69,097502;%%

%\cite{Gross:1991pk}
\bibitem{Gross:1991pk}
  F.~Gross and J.~Milana,
  %``Decoupling Confinement And Chiral Symmetry Breaking: An Explicit Model,''
  Phys.\ Rev.\  D {\bf 45}, 969 (1992).
  %%CITATION = PHRVA,D45,969;%%

%\cite{Engels:2005te}
\bibitem{Engels:2005te}
  J.~Engels, S.~Holtmann and T.~Schulze,
  %``Scaling and Goldstone effects in a QCD with two flavours of adjoint
  %quarks,''
  Nucl.\ Phys.\  B {\bf 724}, 357 (2005)
  [arXiv:hep-lat/0505008].
  %%CITATION = NUPHA,B724,357;%%

%\cite{Kogut:1984sb}
\bibitem{Kogut:1984sb}
  J.~B.~Kogut, J.~Shigemitsu and D.~K.~Sinclair,
  %``Chiral Symmetry Breaking With Octet And Sextet Quarks,''
  Phys.\ Lett.\  B {\bf 145}, 239 (1984).
  %%CITATION = PHLTA,B145,239;%%

%\cite{Banks:1979yr}
\bibitem{Banks:1979yr}
  T.~Banks and A.~Casher,
  %``Chiral Symmetry Breaking In Confining Theories,''
  Nucl.\ Phys.\  B {\bf 169}, 103 (1980).
  %%CITATION = NUPHA,B169,103;%%

%\cite{Leutwyler:1992yt}
\bibitem{Leutwyler:1992yt}
  H.~Leutwyler and A.~Smilga,
  %``Spectrum of Dirac operator and role of winding number in QCD,''
  Phys.\ Rev.\  D {\bf 46}, 5607 (1992).
  %%CITATION = PHRVA,D46,5607;%%

%\cite{Gross:1974jv}
\bibitem{Gross:1974jv}
  D.~J.~Gross and A.~Neveu,
  %``Dynamical Symmetry Breaking In Asymptotically Free Field Theories,''
  Phys.\ Rev.\  D {\bf 10}, 3235 (1974).
  %%CITATION = PHRVA,D10,3235;%%

%\cite{Nambu:1961tp}
\bibitem{Nambu:1961tp}
  Y.~Nambu and G.~Jona-Lasinio,
  %``Dynamical model of elementary particles based on an analogy with
  %superconductivity. I,''
  Phys.\ Rev.\  {\bf 122}, 345 (1961).
  %%CITATION = PHRVA,122,345;%%

%\cite{Nambu:1961fr}
\bibitem{Nambu:1961fr}
  Y.~Nambu and G.~Jona-Lasinio,
  %``Dynamical model of elementary particles based on an analogy with
  %superconductivity. II,''
  Phys.\ Rev.\  {\bf 124}, 246 (1961).
  %%CITATION = PHRVA,124,246;%%

%\cite{Kogut:1998rg}
\bibitem{Kogut:1998rg}
  J.~B.~Kogut, J.~F.~Laga\"{e} and D.~K.~Sinclair,
  %``Thermodynamics of lattice QCD with chiral 4-fermion interactions,''
  Phys.\ Rev.\  D {\bf 58}, 034504 (1998)
  [arXiv:hep-lat/9801019].
  %%CITATION = PHRVA,D58,034504;%%

%\cite{Kogut:2002rw}
\bibitem{Kogut:2002rw}
  J.~B.~Kogut and D.~K.~Sinclair,
  %``The N(t) = 4 finite temperature phase transition for lattice QCD with a
  %weak chiral 4-fermion interaction,''
  arXiv:hep-lat/0211008.
  %%CITATION = HEP-LAT/0211008;%%

%\cite{Clark:2006wp}
\bibitem{Clark:2006wp}
  M.~A.~Clark and A.~D.~Kennedy,
  %``Accelerating Staggered Fermion Dynamics with the Rational Hybrid Monte
  %Carlo (RHMC) Algorithm,''
  Phys.\ Rev.\  D {\bf 75}, 011502 (2007)
  [arXiv:hep-lat/0610047].
  %%CITATION = PHRVA,D75,011502;%%

%\cite{Kogut:2006jg}
\bibitem{Kogut:2006jg}
  J.~B.~Kogut and D.~K.~Sinclair,
  %``The RHMC algorithm for theories with unknown spectral bounds,''
  Phys.\ Rev.\  D {\bf 74}, 114505 (2006)
  [arXiv:hep-lat/0608017].
  %%CITATION = PHRVA,D74,114505;%%

%\cite{Ferrenberg:1988yz}
\bibitem{Ferrenberg:1988yz}
  A.~M.~Ferrenberg and R.~H.~Swendsen,
  %``NEW MONTE CARLO TECHNIQUE FOR STUDYING PHASE TRANSITIONS,''
  Phys.\ Rev.\ Lett.\  {\bf 61}, 2635 (1988).
  %%CITATION = PRLTA,61,2635;%%

%\cite{Bernard:1999fv}
\bibitem{Bernard:1999fv}
  C.~W.~Bernard {\it et al.},
  %``Critical behavior in N(t) = 4 staggered fermion thermodynamics,''
  Phys.\ Rev.\  D {\bf 61}, 054503 (2000)
  [arXiv:hep-lat/9908008].
  %%CITATION = PHRVA,D61,054503;%%

%\cite{D'Elia:2005bv}
\bibitem{D'Elia:2005bv}
  M.~D'Elia, A.~Di Giacomo and C.~Pica,
  %``Two flavor QCD and confinement,''
  Phys.\ Rev.\  D {\bf 72}, 114510 (2005)
  [arXiv:hep-lat/0503030].
  %%CITATION = PHRVA,D72,114510;%%

%\cite{Brown:1988qe}
\bibitem{Brown:1988qe}
  F.~R.~Brown, N.~H.~Christ, Y.~F.~Deng, M.~S.~Gao and T.~J.~Woch,
  %``Nature of the Deconfining Phase Transition in SU(3) Lattice Gauge Theory,''
  Phys.\ Rev.\ Lett.\  {\bf 61}, 2058 (1988).
  %%CITATION = PRLTA,61,2058;%%

%\cite{Bacilieri:1988yq}
\bibitem{Bacilieri:1988yq}
  P.~Bacilieri {\it et al.},
  %``ON THE ORDER OF THE DECONFINING PHASE TRANSITION IN PURE GAUGE QCD,''
  Phys.\ Rev.\ Lett.\  {\bf 61}, 1545 (1988).
  %%CITATION = PRLTA,61,1545;%%

\end{thebibliography}
\end{document}